\documentclass[aps,floats,showpacs,amsmath,epsf,twocolumn, prx]{revtex4-1}
\usepackage{amsfonts, relsize, color}
\usepackage{graphics}
\usepackage{graphicx}
\usepackage{hyperref}

\begin{document}
\title{Excitonic and Nematic Instabilities on the Surface of Topological Kondo Insulators}

\author{Bitan Roy}
\email{Corresponding author: broy@umd.edu}
\affiliation{Condensed Matter Theory Center and Joint Quantum Institute, University of Maryland, College Park, Maryland 20742-4111, USA}

\author{Johannes Hofmann}
\affiliation{Condensed Matter Theory Center and Joint Quantum Institute, University of Maryland, College Park, Maryland 20742-4111, USA}

\author{Valentin Stanev}
\affiliation{Condensed Matter Theory Center and Joint Quantum Institute, University of Maryland, College Park, Maryland 20742-4111, USA}

\author{Jay D. Sau}
\affiliation{Condensed Matter Theory Center and Joint Quantum Institute, University of Maryland, College Park, Maryland 20742-4111, USA}

\author{Victor Galitski}
\affiliation{Condensed Matter Theory Center and Joint Quantum Institute, University of Maryland, College Park, Maryland 20742-4111, USA}
\affiliation{School of Physics, Monash University, Melbourne, Victoria 3800, Australia}

\date{\today}

\begin{abstract}
We study the effects of strong electron-electron interactions on the surface of cubic topological Kondo insulators (such as samarium hexaboride, SmB$_6$). Cubic topological Kondo insulators generally support three copies of massless Dirac nodes on the surface, but only two of them are energetically degenerate and exhibit an energy offset relative to the third one. With a tunable chemical potential, when the surface states host electron and hole pockets of comparable size, strong interactions may drive this system into rotational symmetry breaking nematic and translational symmetric breaking excitonic spin- or charge-density-wave phases, depending on the relative chirality of the Dirac cones. Taking a realistic surface band structure into account we analyze the associated Ginzburg-Landau theory and compute the mean field phase diagram for interacting surface states. Beyond mean field theory, this system can be described by a two-component isotropic Ashkin-Teller model at finite temperature, and we outline the phase diagram of this model. Our theory provides a possible explanation of recent measurements which detect a two-fold symmetric magnetoresistance and an upturn in surface resistivity with tunable gate voltage in SmB$_6$. Our discussion can also be germane to other cubic topological insulators, such as ytterbium hexaboride (YbB$_6$), plutonium hexaboride (PuB$_6$).
\end{abstract}

\pacs{ 73.20.-r, 71.35.Lk}

\maketitle

\vspace{10pt}

\section{Introduction}

 It was realized in the past decade that the band structure of a strongly spin-orbit coupled three-dimensional solid with preserved time-reversal and inversion symmetries can be associated with a topological Z$_2$ index~\cite{review-1, review-2}. A system with such nontrivial topological index, also known as strong $Z_2$ topological insulator, belongs to class AII in ten fold way of classification~\cite{tenfold-3d}. These materials ideally have an insulating bulk but host an odd number of metallic surface states which are protected against time-reversal invariant perturbations. Typical topological insulators (such as Bi$_2$Se$_3$) are often only very weakly correlated. Our theoretical understanding of these materials is thus based on a noninteracting electronic band structure picture that is not affected by the presence of weak electron-electron interactions. Within the same class (AII), a \emph{strongly correlated} topological Kondo insulator (TKI) was predicted to exist in Ref.~\cite{coleman-gatilski-dzero-sun}, in which the hybridization between localized $f$- and conduction $d$-electrons opens up a topologically nontrivial bulk-insulating gap below the Kondo temperature. Indeed, a number of recent experiments are strongly suggesting that samarium hexaboride (SmB${}_6$) possibly supports a TKI below the Kondo temperature (~50 K)~\cite{exp1, exp2, exp3, exp4, exp5, exp6, exp7, exp8, exp9, xia-fisk-NatMat}. The bulk topological invariant can be computed within the mean-field description of this system, yielding a nonzero $Z_2$ index. These recent findings motivate the search for effects where both interactions and topological details play crucial role at low temperatures~\cite{efimkin-galitski, nikolic, hridis}.

Motivated by the possibility that TKIs can be a fertile ground to support novel interplay of topology and correlations, we here consider the effect of strong electronic interactions on the surface of TKIs and demonstrate that gapless surface states in these systems can be susceptible towards nematic and excitonic density-wave phases. We also show that our theoretical analysis can be germane to two recent experiments~\cite{c4-magnetiresis,paglione}, which could be indicative of interaction-induced instabilities on the surface of a TKI: first, a magnetoresistance measurements on SmB${}_6$ reports a C${}_2$ and C${}_4$-symmetric magnetoresistance at low and high temperatures, respectively~\cite{c4-magnetiresis}. These findings indicate a rotational symmetry breaking nematic ordering on the surface of a TKI. Second, Ref.~\cite{paglione} reports a measurement of the surface resistivity in SmB$_6$ where the resistivity increases with varying gate voltage, which may, for example, arise due to an underlying excitonic ordering. In this work we develop a theory for the interacting surface states in TKIs, which provides possible explanations to these observations.

\begin{figure}[t]
\includegraphics[width=8.00cm,height=6.50cm]{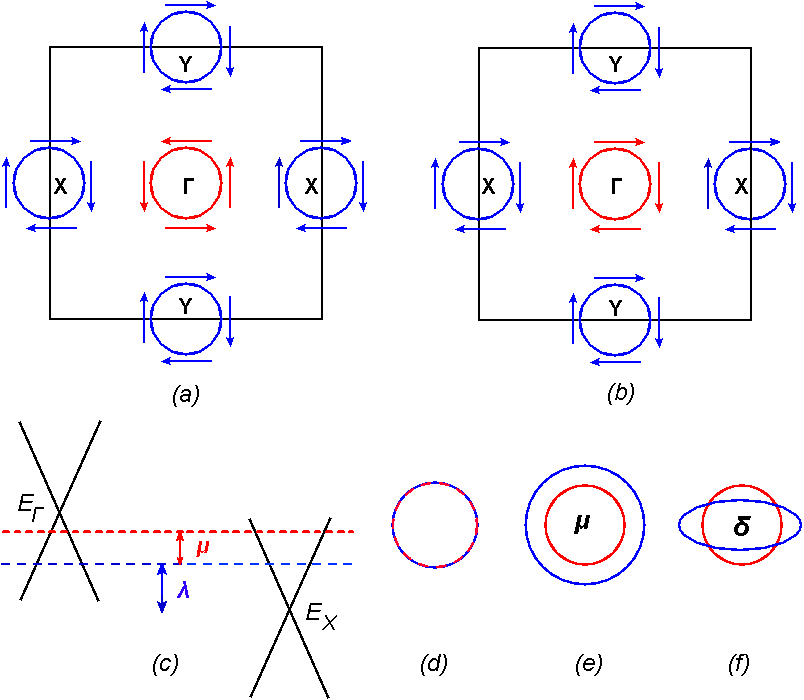}
\caption[] {(Color online) Top row: Two possible chiralities of electron (blue) and hole (red) pockets on the surface of cubic TKIs, leading to an excitonic instability in the (a) SDW and (b) CDW channels, respectively. (c) Offset among the Dirac points near the $\Gamma$ and $X/Y$ points, and (d) deviation from perfectly nesting, due to (e) unequal sizes of the pockets, (f) ellipticity in the electron pocket, parametrized by $\mu$ and $\delta$, respectively. }\label{fermisurfacetopology}
\end{figure}

Consider the typical surface band structure of a \emph{cubic} topological insulator (for example, SmB${}_6$): these systems are strong $Z_2$ topological insulators and thus support an odd number of metallic surface states. In the cubic environment of SmB$_6$, the band inversion takes place at the three $X$ points of the bulk Brillouin zone (BZ) \cite{coleman-alexandrov, takimoto}. Hence, an interface of a cubic TI with the vacuum supports three copies of massless Dirac cones at the $\Gamma$, $X$, and $Y$ points of the surface BZ, as illustrated in Figs.~\ref{fermisurfacetopology}(a) and (b) [throughout the paper, we assume that the surface is cleaved along a high symmetry axis, such as $(001)$]. The underlying cubic symmetry enforces equal energies $E_X$ and $E_Y$ of the Dirac nodes at the $X$ and $Y$ points, respectively, which manifests a four-fold rotational $C_4$ symmetry on the surface. The $\Gamma$ Dirac point is, however, not constrained by this symmetry and generically displays an offset with respect to the $X$ and $Y$ points, i.e., $E_\Gamma \neq E_{X/Y}$ (we set $E_\Gamma> E_{X/Y}$ in the remainder to be definite), which can be as large as $\sim 10-12$ meV \cite{roy-sau-dzero-galitski, xidai}. This surface band structure is also in agreement with recent ARPES measurements~\cite{exp4, exp5, exp6, exp7, exp8, exp9, ybb6-1, ybb6-2, ybb6-3}. Due to such large energy off-set among the Dirac points, it is natural to anticipate that surface chemical potential is tuned in between $E_\Gamma$ and $E_{X,Y}$, giving rise to electron and hole pockets that can be conducive for excitonic condensation. If, on the other hand, all the pockets are electron or hole like such configuration can be achieved through external gating, for example~\cite{paglione}.

It is therefore conceivable to place the chemical potential in between $E_\Gamma$ and $E_{X/Y}$ \cite{paglione}, yielding one hole pocket around the $\Gamma$ point and two electron pockets near the $X$, $Y$ points, as shown in Fig.~\ref{fermisurfacetopology}(c). Now, if interactions on the surface are included, electrons in the $X/Y$ pockets can pair via the so called the Keldysh-Kopaev mechanism with holes in the $\Gamma$ pockets~\cite{keldysh-kopaev}, giving rise to an excitonic condensate i.e., a density wave, which is modulated by half the reciprocal lattice vector of the surface BZ. This paper discusses the phase diagram of this effective interacting surface theory.

Since, the underlying bulk theory is strongly spin-orbit coupled, spin (planar components) and momentum of the surface Dirac cones will be locked as shown in Fig.~\ref{fermisurfacetopology}(a) and (b). Therefore, only the $z$-component of the spin remains free and participates in the ordering. However, in principle, two distinct possible types of excitonic instabilities can occur on the surface of TKIs depending on the relative chirality of the Dirac cones at the $X/Y$ and $\Gamma$ points. When all Dirac cones on the surface have identical chirality [Fig.~\ref{fermisurfacetopology}(a)], the excitonic condensate is formed by electrons and holes with opposite spin projection, giving rise to triplet spin-density wave (SDW) order. If, on the other hand, the Dirac points at the $\Gamma$ and $X,Y$ points carry opposite chirality [Fig.~\ref{fermisurfacetopology}(b)], pairing occurs between particles and holes with equal spin projection, leading to singlet charge-density wave (CDW) order. Here singlet and triplet orders are defined in terms of total angular momentum. Our discussion is, however, insensitive to the exact nature of the excitonic ordering, and we thus assume equal chirality for all Dirac cones and discuss the SDW instability in the following.

Currently there is an ongoing debate on the effective model for bulk insulating state in SmB$_6$ that can lead to different spin texture on the surface~\cite{xidai, vojta}. However, the nature of the excitonic order only depends on the relative chirality of electron- and hole-like Dirac surfaces. Recent theoretical works~\cite{sigrist, vojta-2} have demonstrated that depending on the relative strength of nearest-neighbor and next-nearest-neighbor hybridization among $d$- and $f$-electrons, one can realize either two scenarios, we presented in Fig.~\ref{fermisurfacetopology}. Therefore, our classification exhausts all possibilities for the excitonic order and the following discussion is insensitive to the details of the bulk band structures (since, the SDW and the CDW orders give identical phase diagram).

If the Fermi surfaces are perfectly nested [as shown in Fig.~~\ref{fermisurfacetopology} (d)], the Keldysh-Kopaev mechanism dictates that an excitonic instability sets in for arbitrarily weak repulsive interactions. It turns out, however, that a realistic surface band structure deviates from perfect nesting in two ways: first, generically the chemical potential will not be exactly placed in the middle between $E_\Gamma$ and $E_{X/Y}$ [as illustrated in Fig.~\ref{fermisurfacetopology} (e)]. This Fermi surface mismatch reduces the propensity for excitonic pairing, analogous to the Clogstron-Chandrasekhar effect in standard BCS theory where the chemical potential imbalance is induced by a Zeeman term. Second, recent band structure calculations~\cite{roy-sau-dzero-galitski} indicate that only the $\Gamma$ Dirac cone is isotropic while both $X$ and $Y$ Dirac cones can be anisotropic [see Fig.~\ref{fermisurfacetopology} (f)], in agreement with ARPES measurements~\cite{exp4, exp5, exp6, exp7, exp8, exp9, ybb6-1, ybb6-2, ybb6-3}. We take these realistic effects into account, finding that the overall structure of the phase diagram is not strongly affected by these effects, although they may reduce the transition temperature of various orderings.

We note that the surface band-structure shown on Fig.~\ref{fermisurfacetopology} exhibits strong resemblance to the band structure of the iron-based superconductors~\cite{chubukov-metallic}. We discuss both the similarities and the differences between these systems at the end of the paper (see Sec.~VI). The excitonic ordering due to weak repulsive interactions, known as Keldysh-Kopaev mechanism~\cite{keldysh-kopaev}, has also been exploited to address the SDW instability in Cr \cite{rice} and iron-based superconductors \cite{czetkovic-tesanovic, Chubukov,Han}, antiferromagnetic ordering for weak Hubbard repulsion in monolayer \cite{aleiner-kharzeev-tsvelik, assaad-graphene} and bilayer \cite{roy-yang} graphene, 2D Kondo insulators \cite{assaad-kondo} when placed in an in-plane magnetic field, and in the context of possible excitonic instability in topologically trivial calcium hexaboride (CaB$_6$) \cite{cab6}.

This paper is structured as follows: in Sec.~\ref{sec:microscopics}, we introduce the microscopic description of the interacting surface states of a cubic TKI. In Sec.~\ref{sec:ginzburglandau}, we discuss the Ginzburg-Landau theory of the model that is valid in the vicinity of a second order phase transition at finite temperature. In particular, we discover that in the limit of small ellipticity the order developing on the surface breaks the $C_4$ lattice symmetry down to $C_2$. We further illustrate that the condensation of excitonic order parameters only breaks discrete symmetries, thus implying that true long range order is described accurately by a mean field analysis of our effective surface theory. We present a full numerical computation of the mean field phase diagram in Sec.~\ref{sec:meanfield}, finding a second order phase transition for nearly perfect nesting from a high-temperature paramagnetic phase to a $C_2$-symmetric state at low temperature in which an excitonic condensate develops between the $\Gamma$ and the either $X$ or $Y$ pockets, but not both, thus spontaneously breaking the discrete $C_4$ symmetry of the surface BZ. As is well known, mean field theory does not assume correlations in the paramagnetic or normal phase at high temperature and does not distinguish between a phase where true long range order develops in the form of a nematic phase (with broken rotational symmetry) where thermal fluctuations dominate~\cite{fradkin} and a density wave phase  (with broken translational symmetry). These two distinct transitions, which coincide in mean field theory, can, in principle, take place at different temperatures. This occurs through the proliferation of domain walls in the system. It turns out that the effective theory describing the dynamics of the domain walls can be mapped onto a two-component isotropic \emph{Ashkin-Teller} model. We exploit such mapping in Sec.~\ref{sec:ashkinteller} to elucidate the phase diagram beyond the mean field level. Finally, the paper is concluded by a summary and discussion in Sec.~\ref{sec:summary}. In particular, we comment on similarities as well as some differences between our findings and the phase structure in a completely different class of systems, the iron pnictides.

\section{Microscopic Hamiltonian for interacting surface states}\label{sec:microscopics}

This section introduces the microscopic description of the interacting surface states. The appropriate spinor basis is chosen to be $\Psi_j=\left( \Psi_{\uparrow,j}, \Psi_{\downarrow,j} \right)$, near $j=\Gamma, X, Y$ points of the surface BZ, where $\Psi_{\sigma,j}$ is composed of linear superposition of $d$ and $f$ electrons with spin projection $\sigma=\uparrow, \downarrow$. The relative weight among $d$ and $f$ electrons in the surface states is set by the bulk band parameters, such as hopping amplitudes and hybridization matrix elements \cite{roy-sau-dzero-galitski}. A recent transport measurements in SmB$_6$ with different thickness clearly establish that a low temperatures (sufficiently below the Kondo temperature) surface states are decoupled from the bulk and the transport properties are essentially determined by the former ones \cite{exp2}. Furthermore, spin-resolved ARPES has established the helical spin-texture of the surface states, and quantum oscillation has observed the signature of Dirac Landau levels up to 45 Teslas~\cite{luli}. Also, recent thermo-electric measurements captured the signature of heavy Dirac fermions on the surface, even after mechanically damaging the surface ~\cite{thompson}. These observations strongly indicate that despite small bulk gap ($\sim 15$ meV), and large number of bulk states, the surface and the bulk states are effectively decoupled in SmB$_6$ that in turn allows us to treat the gapless surface states separately~\cite{commets-bulkins}. Notice, in YbB$_6$ the bulk gap is $\sim 100$ meV and one can safely neglect any coupling between bulk and surface states.

The noninteracting Hamiltonians describing the helical Dirac fermionic excitations near the $\Gamma,X$ and $Y$ points take the form  [setting $\hbar=1$] 
\begin{equation}\label{freeHamil}
H_j= v^j_x k_x \sigma_x-v^j_y k_y \sigma_y,
\end{equation} 
where $j=\Gamma, X, Y$, with $v^\Gamma_x=v^\Gamma_y=v$ as the Fermi velocity of the isotropic Dirac cone near the $\Gamma$ point. The underlying C$_4$ symmetry of the surface BZ implies $v^X_x=v^Y_y$ and $v^X_y=v^Y_x$. The ellipticity of the Dirac cones near the $X$ and $Y$ points is captured by defining $v^X_x=v(1+\delta)$ and $v^X_y=v(1-\delta)$. The parameter $\delta$ in SmB$_6$ ranges from $0.1$ to $0.4$ \cite{exp4, exp7, exp8, exp9}. The above form of the Hamiltonian is restricted by the bulk topological invariant and in momentum-space they represent \emph{anti-vortices} near $\Gamma, X,Y$ points of the surface BZ, capturing the signature of nontrivial topological invariant of the bulk insulating state on the surface.

Excitonic SDW ordering arises from a repulsive interaction between fermions with opposite spin projections in the $\Gamma$ and $X,Y$ pockets. Such a particle-hole pairing instability can be taken into account by adding a repulsive short-ranged interaction
\begin{equation}
H_{\rm int} = - \frac{U_0}{2} \sum_{j=X,Y} \int \frac{d^2q}{(2\pi)^2} \, s_{j,{\bf q}}^\dagger s_{j,{\bf q}}^{} \label{eq:Hint} 
\end{equation}
to the free Hamiltonian ($H_j$), where $U_0>0$ and
\begin{equation}
s_{j,{\bf q}} = \int \frac{d^2k}{(2\pi)^2} \, c_{\Gamma,{\bf k}+{\bf q}\alpha}^\dagger (\sigma_3)_{\alpha\beta}^{} c_{j,{\bf k}\beta} \label{eq:sqdef}
\end{equation}
is the spin operator. $c_{j,{\bf k}\alpha}^\dagger$ creates a fermion in the $j=\Gamma,X,Y$ pocket with momentum ${\bf k}$ and spin $\alpha$. The momentum of the $X$ and $Y$ excitation is measured relative to the nesting vectors ${\bf Q}_X=(\pi,0)$ and ${\bf Q}_Y=(0,\pi)$. In Eq.~\eqref{eq:sqdef}, a summation over the spinor indices $\alpha$ and $\beta$ is implied. Within the same framework, CDW ordering can be studied by simply replacing the Pauli matrix $\sigma_3$ by $\sigma_0$ in Eq.~(\ref{eq:sqdef}) and changing the sign of one matrix in $H_{X/Y}$ or $H_\Gamma$ in Eq.~(\ref{freeHamil}), without quantitatively changing the results. The order parameter for the excitonic SDW condensation is
\begin{align}
\Delta_{X/Y} = \frac{U_0}{2} \langle c_{\Gamma,{\bf k}\alpha}^\dagger (\sigma_3)_{\alpha\beta} c_{X/Y,{\bf k}\beta}^{}\rangle , \label{eq:orderparameter}
\end{align}
where $\langle \ldots \rangle$ denotes the thermal expectation value.

\section{Ginzburg-Landau Theory}\label{sec:ginzburglandau}

 In this section, we discuss the Ginzburg-Landau expansion of the ordered state, which describes the second order phase transition at small $\delta$. This analysis will allow us to gain a qualitative insight into the phase diagram for interacting surface states of TKIs and the notion of symmetry breaking in various ordered phases. The Ginzburg-Landau functional can be constructed by systematically expanding the free energy ${\cal F}$ in powers of $\Delta_X$ and $\Delta_Y$, yielding
\begin{eqnarray}
 \mathcal{F}(\Delta_i) &=&  \mathcal{K}\left[ (|\vec{\nabla} \Delta_{X}|)^2  + (|\vec{\nabla} \Delta_{Y}|)^2 \right] + \alpha [ |\Delta_{X}|^2 + |\Delta_{Y}|^2] \nonumber\\ 
&+& \frac{\beta}{2}(|\Delta_{X}|^2  +|\Delta_{Y}|^2)^2 + \gamma |\Delta_{X}|^2|\Delta_{Y}|^2.
\label{GL1}
\end{eqnarray}
The last term (proportional to $\gamma$) plays an important role in determining the pattern of symmetry breaking in the ordered phase. 
For $\gamma=0$, the free energy is degenerate for fixed $|\Delta_X|^2 + |\Delta_Y|^2$. If $\gamma >0$, surface states develop a finite expectation value of either $|\Delta_X|$ or $|\Delta_Y|$, but not both. Such a phase manifestly breaks the C$_4$ rotational symmetry down to C$_2$, and the system simultaneously develops a nematic order. On the other hand, when $\gamma <0$, the system minimizes the free energy by simultaneously condensing $|\Delta_X|$ and $|\Delta_Y|$ at the same temperature, and the four-fold C$_4$ rotational symmetry of the system is preserved in the ordered phase.

In terms of the microscopic parameters, $\gamma$ reads~\cite{fernandez-chubukov-schmalian}
\begin{eqnarray}
 \gamma = \text{\bf{Tr}}[\hat{G}_{\Gamma} \hat{G}_X \hat{G}_{\Gamma} \hat{G}_X + \hat{G}_{\Gamma} \hat{G}_Y \hat{G}_{\Gamma} \hat{G}_Y -2 \hat{G}_{\Gamma} \hat{G}_X \hat{G}_{\Gamma} \hat{G}_Y],
\label{gamma2}
\end{eqnarray}
where we define
\begin{equation}
\hat{G}^{-1}_{\Gamma}= -i \omega + H_\Gamma -\lambda_-, 
\hat{G}^{-1}_{X/Y}= -i \omega + H_{X/Y} +\lambda_+ ,
\end{equation}
with $\lambda_\pm = \lambda \pm \mu$ [see also Fig.~\ref{fermisurfacetopology}(c)] and  $\text{\bf{Tr}}$ implies a summation over momentum, Matsubara frequency, and spinor indices. If all bands are perfectly circular ($\delta=0$), $\hat{G}_X = \hat{G}_{Y}$ and concomitantly $\gamma=0$, which remains true even if the bands are not perfectly nested, i.e., $\mu \neq 0$. In a realistic situation with elliptic eletron-like Fermi pockets near $X$ and $Y$ points (i.e., $\delta \neq 0$), we have $\gamma \neq 0$. For small ellipticity ($\delta \ll 1$), expanding all the quartic terms in ${\cal F}$ in powers of $\delta$, we obtain $\gamma=\delta^2 g(T, \mu)$, where $g(T_c,\mu)$ is a {\it positive} function close to $T_c$. Thus, the SDW state breaks the $C_4$ symmetry on the surface under $X\leftrightarrow Y$. In the limit of large ellipticity, we must treat $\delta$ non-perturbatively, which is done in the next section.

We point out that the Ginzburg-Landau functional in Eq.~\eqref{GL1} possesses a $U(1)$ valley symmetry of the SDW OPs ($\Delta_X$, $\Delta_Y$) associated with their phases $\Delta_j=|\Delta_j| e^{i \phi_j}$, which implies that for $\gamma>0$ the SDW order not only spontaneously breaks a discrete C$_4$ rotational but also a continuous U$(1)$ symmetry. It is important to note that such continuous $U(1)$ symmetry is only an artifact of the low energy approximation for the surface states and can be reduced if we allow an additional quartic term
\begin{align}
\mathcal{F}_{SB}=\rho |\Delta_X|^2 |\Delta_Y|^2 \left[ \cos(2 \varphi_X)+ \cos(2 \varphi_Y)\right]
\label{SymBreaking}
\end{align}
 in Eq. (\ref{GL1}). Such a term can, for example, be generated by pair-scattering processes represented by $c_{\Gamma}^{\dagger} c_{\Gamma}^{\dagger} c_{X} c_{X}$ and $c_{\Gamma}^{\dagger} c_{\Gamma}^{\dagger} c_{Y} c_{Y}$, also known as Umklapp processes, which are allowed in the presence of an underlying lattice \cite{roy-herbut-kekule}. The physical origin of such terms can be appreciated in the following way: the phase degree of freedom of $\Delta_j$ represents a \emph{sliding mode} of the SDW order in real space. However, in any material the commensurate density wave will be pinned to the lattice.  Hence, we need to take into account such lattice-induced terms to pin density-wave order, that also reduce the (artificial) valley U$(1)$ symmetry down to a discrete $Z_2$ one. Most importantly, this implies that no continuous symmetry is broken and the SDW order on the two-dimensional surface of cubic TKIs can exhibit true long-range order~\cite{marmin-wagner}. In particular, we expect that a mean field analysis provides an accurate phase diagram of the effective surface theory, despite the fact that the system is two-dimensional. We discuss the mean field phase diagram in the next section.

\section{Mean Field Phase Diagram}\label{sec:meanfield}

To go beyond the Ginzburg-Landau regime of the phase diagram, we now analyze the interacting surface theory in the mean field approximation. In this section we neglect the symmetry-breaking terms [Eq.~(\ref{SymBreaking})], and thus the excitonic orders enjoy an artificial U$(1)$ symmetry. In terms of the order parameters, defined in Eq.~\eqref{eq:orderparameter}, the free energy density reads
\begin{equation}
{\cal F} =\frac{2}{U_0} \bigl(|\Delta_X|^2 + |\Delta_Y|^2\bigr) - \frac{1}{2 \beta} \sum_{i=1}^6 \int \frac{d^2k}{(2\pi)^2}  \ln \left[ 2 \cosh \frac{\beta E_i}{2} \right], \label{eq:freenergy}
\end{equation}
where $\beta$ is the inverse temperature and $E_i$ are the six eigenvalues of the effective quadratic single-particle Hamiltonian
\begin{equation} 
H_{\rm HS} = \left[ 
\begin{array} {ccc}
H_\Gamma -\lambda_- \sigma_0 & - \Delta_X \sigma_3 & - \Delta_Y \sigma_3 \\
- \Delta^\dagger_X \sigma_3 & H_X +\lambda_+ \sigma_0 & 0 \\
- \Delta^\dagger_Y \sigma_3 & 0 & H_Y +\lambda_+ \sigma_0
\end{array}
\right]. \label{eq:HSHamiltonian}
\end{equation}
In the above equation, we set $\lambda_\pm=\lambda \pm \mu$ as illustrated in Fig. 1(a). As is characteristic for two-dimensional Dirac systems, the free energy density in Eq.~(\ref{eq:freenergy}) diverges linearly due to large-momentum contributions, which, however, can be absorbed in a renormalization of the effective interaction strength
\begin{equation}
\frac{1}{U_0} = \frac{1}{U} - \frac{2}{v^2}  \Lambda ,
\end{equation}
where $U>0$ is the renormalized interaction and $\Lambda$ is an ultraviolet cutoff in momentum space \cite{hofmann10,herbut-roy-regularization}, which in real systems corresponds to the bulk band gap. Consequently, physical quantities only depend on $U$ but not on the non-universal cutoff scale $\Lambda$ or the bare coupling $U_0$.

\begin{figure*}[t]
\raisebox{1cm}{\scalebox{0.4}{\includegraphics{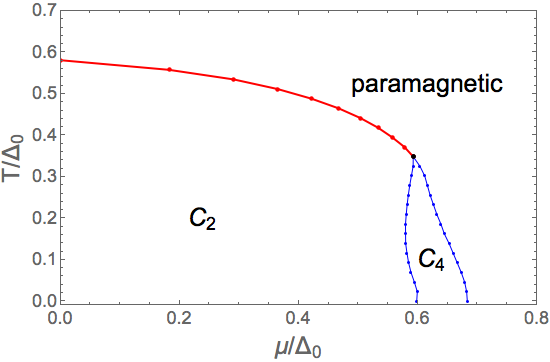}}} 
\scalebox{0.4}{\includegraphics{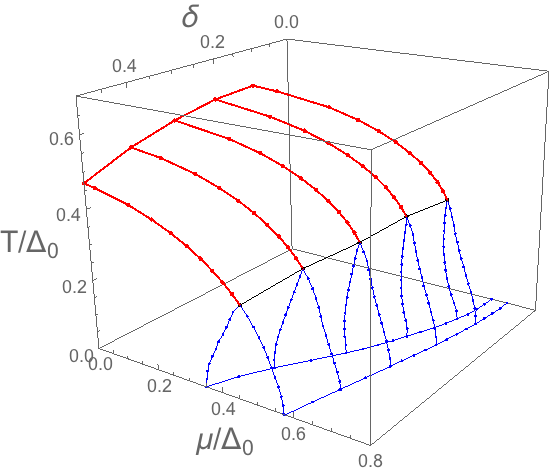}}
\caption[] {(Color online) Left: phase diagram for $\lambda = 2 U^{-1}$ and $\delta = 0.2$ as a function of Fermi surface mismatch $\mu$ and temperature $T$. At small temperature and chemical potential, the ground state has only a $C_2$ symmetry. Red (thick) and blue (thin) lines denote second and first order phase transitions, respectively. Right: Phase diagram for $\lambda = 2 U^{-1}$ as a function of Fermi surface anisotropy $\delta$, $\mu$, and $T$. Notation as in the left panel. The parameter $\delta$ in various ARPES experiments are $\delta=0.25$~\cite{exp4}, $0.11$~\cite{exp7}, $0.21$~\cite{exp8}, $0.33$~\cite{exp9}. It is worth pointing out that the phase diagrams we obtain here are qualitatively similar to the one extracted experimentally for iron pnictides, which also share similar structure of the BZ \cite{Avci}.}\label{fig:phase0p2}
\end{figure*}

In Fig.~\ref{fig:phase0p2} (left), we present the phase diagram as obtained by minimizing the free energy in Eq.~(\ref{eq:freenergy}) as a function of chemical potential $\mu$ and temperature $T$ for the nesting $\lambda = 2 U^{-1}$ and an ellipticity of the $X/Y$ surface pockets of $\delta = 0.2$. At small chemical potential, the ground state displays a two-fold rotational or C$_2$ symmetry, where electrons from either $X$ or $Y$ pocket pair with holes from the $\Gamma$ point, respectively, yielding $|\Delta_X|\neq 0$ and $|\Delta_Y|=0$ or vice versa. The appearance of C$_2$ SDW order naturally introduces a nematicity (characterized by $\Delta_X \neq \Delta_Y$) in the system. As the temperature is increased, there is a continuous second-order transition out of the SDW phase to the paramagnetic (PM) phase.

This limit corresponds to the Ginzburg-Landau analysis presented in the previous section. If the chemical potential (and hence the Fermi surface mismatch) is increased, the direct C$_2$-normal (PM) transition at low temperature is masked by an intermediate phase in which the C$_4$ rotational symmetry is restored and all Fermi pockets participate in the excitonic pairing. Both C$_2$-C$_4$ and C$_4$-PM transitions are first order in nature. Figure~\ref{fig:phase0p2} (right) shows the complete phase diagram as a function of $\mu$, $\delta$, and $T$. Increasing the ellipticity $\delta$ pushes the critical chemical potential for the C$_2$-C$_4$ transition to smaller values but only mildly affects the subsequent C$_4$-PM transition. Hence, while a small ellipticity favors the C$_2$ phase at small $\mu$, the region of the phase diagram with C$_4$-symmetry increases when the Fermi surfaces are strongly anisotropic.

 There is an intuitive picture why at small ellipticity the system is $C_2$ symmetric and only at large Fermi surface mismatch the $C_4$ symmetric phase arises~\cite{brydon}: for nearly perfect nesting (small $\delta$ and $\mu$), the same hole-like state near the $\Gamma$ point contribute to the excitonic pairing with electron-like states from $X$ and $Y$ pockets. Thus, pairing between $\Gamma$ and $X$ reduces the available phase space for pairing between $\Gamma$ and $Y$ and vice versa, implying that only one condensate develops and the system enters the $C_2$ symmetric phase. As the Fermi surface mismatch increases, however, disjoint regions of the $\Gamma$ Fermi surface contribute to the excitonic condensation and a $C_4$ symmetric phase becomes preferable, as demonstrated by our full calculation of the phase diagram.

For $\delta=0$ (circular Fermi surfaces) the quadratic Hamiltonian in Eq.~\eqref{eq:HSHamiltonian} manifests a U(1) symmetry among the exitonic OPs $\Delta_X$ and $\Delta_Y$, and consequently the free energy depends only on the magnitude $\Delta^2 = |\Delta_X|^2+|\Delta_Y|^2$. Thus, in the limit $\delta=0$, there is no distinction between the C$_2$ and the C$_4$ symmetric phases. At zero temperature, the free energy density then takes the particularly simple form ${\cal F} = \mu^2 - \Delta_0^2/2$, where $\Delta_0$ represents the SDW OP at $T=0$ and $\mu=0$, which implies a first-order transition between condensed and normal phase at the standard critical Clogston-Chandrasekhar value $\mu_{\rm crit}=\Delta_0/\sqrt{2}$, which can also be seen in Fig. \ref{fig:phase0p2} (right).

We point out that the structure of the BZ in iron-based superconductors is qualitatively similar to the one for the surface states of cubic TKIs. Interestingly, the phase diagrams of these two systems bear some qualitative similarities \cite{chubukov-metallic,brydon, Avci}. In particular, the C$_2$-C$_4$ phase transition that can be tuned by doping has been observed experimentally in pnictide materials~\cite{Avci}.

 We note that there are two possible ways to modify the mean field phase diagram: for a large Fermi surface anisotropy, the system may condense into an incommensurate density-wave phase, where the periodicity of the excitonic condensate is different from the reciprocal lattice vector~\cite{rice}. Furthermore, for large doping, various superconducting instabilities may set in. The discussion of these phenomena is beyond the scope of this paper.

Our present mean field analysis does not account for thermal fluctuations. Quite generally, a full analysis of the phase diagram should, in principle, distinguish between the nematic and the excitonic phases. As will be discussed in the following section, once thermal fluctuations are incorporated, the transition temperatures for these two instabilities can be different. Let us focus on the regime of small chemical potential, where mean-field theory predicts a $C_2$ phase for arbitrary $\delta$, as shown in Fig~\ref{fig:phase0p2}. In this phase either $\Delta_X$ or $\Delta_Y$ develops a nonzero but real expectation value and thus the surface states simultaneously develop a \emph{nematic} (due to the breaking of $C_4$ symmetry) as well as a translational symmetry breaking commensurate SDW order. These orders can be represented by two different Ising-like variables and thus the ground state at $T=0$ displays an exact four-fold degeneracy. However, at finite temperature, thermal fluctuations allow the system to fragment into multiple domains of these degenerate phases. We we will argue that interplay of these domains at finite temperatures can be captured by a two-component isotropic Ashkin-Teller model, and allude to the finite temperature phase diagram for the surface states beyond the mean field approximation.

Before concluding the section a discussion on the nature of the nematic order seems appropriate. Notice that the nematic phase is described by a fluctuating excitonic order that does not acquire a finite vacuum expectation value. As pointed out in the Introduction that depending on the relative strength of nearest-neighbor and next-nearest-neighbor hybridization amplitude among the opposite parity orbitals (such as $d$ and $f$) in the bulk, the chiralities of electron and hole pockets can be same or opposite, which in turn determines the nature of density-wave excitonic order (SDW or CDW). Therefore, depending on bulk hybridization strength over a finite range, the nematic phase may represent either a fluctuating charge- or spin-density-wave order. However, the phase diagram of the interacting surface states is insensitive to the exact nature of the ordering, as only discerte Ising-like symmetries are broken in the charge- or spin-density-wave phases (uniform or fluctuating).

\section{Thermal fluctuations, domain walls and Ashkin-Teller model}\label{sec:ashkinteller}

To understand the role of a domain walls at finite temperatures, we first consider a simpler situation, where the system exhibits only a two-fold degeneracy among the configurations, say $A$ and $B$ [chosen from four possible states with $\Delta_X > 0$ or $\Delta_X<0$ and $\Delta_Y>0$ or $\Delta_Y<0$]. The free energy of the domain-wall per unit length of this system is given by $F=J_{AB}-T S_{AB}$, where $S_{AB} (J_{AB})$ is the entropy (energy) per unit length of a single domain wall. For temperatures $T>J_{AB}/S_{AB}$, we have $F<0$, and the free-energy is minimized through the proliferation of domain walls between these two configurations.

\begin{figure}[t]
\includegraphics*[width=7.50cm,height=5.25cm]{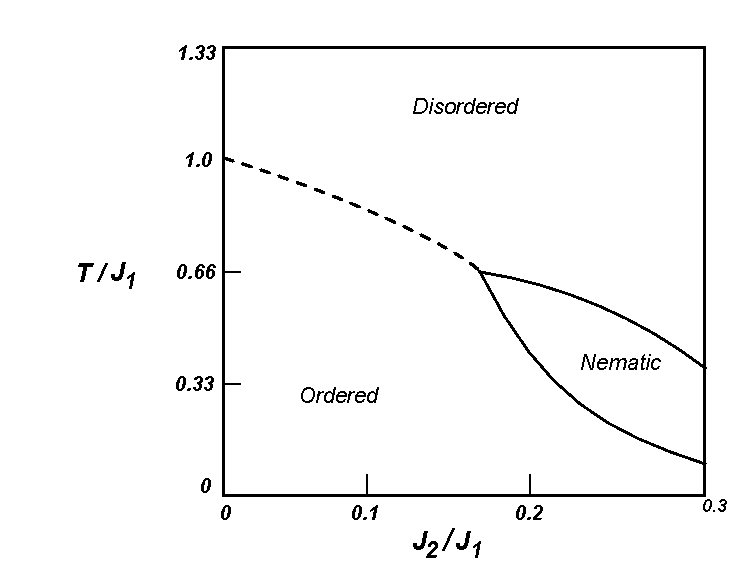}
\caption[] { Phase diagram of the two-component isotropic Ashkin-Teller Model \cite{kadanoff-etal}. In terms of microscopic parameters $J_1 \sim U$ and $J_2/J_1 \sim \delta^2$.}\label{ashkinteller}
\end{figure}

To estimate the result of proliferation of domain walls on the surface of cubic TKIs, we define two Ising-spin variables $s=\textrm{sgn}(|\Delta_X|-|\Delta_Y|)$ and $\sigma=\textrm{sgn}(\Delta_X+\Delta_Y)$. The spin variable $s$ determines the direction of the SDW order, while $\sigma$ represents how the translation symmetry is broken. Therefore, in the nematic phase $s \neq 0$, and when the density-wave order condenses we have $\sigma \neq 0$. The energy of the domain walls can be accounted for by an effective exchange Hamiltonian
\begin{equation}\label{exch}
H_{ex}=-\sum_{\langle i,j \rangle} \left[ J_{2} \; s_i s_j + J_{1}(1+s_i s_j)\sigma_i \sigma_j \right],
\end{equation}
where $J_{2}$ represents the energy a domain wall between the regions where $|\Delta_X|\neq 0$ and $|\Delta_Y|\neq 0$. $J_{1}$ represents a similar quantity where $\Delta_X$ or $\Delta_Y$ changes the sign without changing the direction of the symmetry breaking (hence the factor $(1+s_i s_j)$). We expect $J_{2}/J_{1}$ to be proportional to $\delta^2$, where $\delta$ is the ellipticity of the pockets near $X$ and $Y$ points. In terms of a redefined variable $s\rightarrow \tilde{s}= s\sigma$, the rescaled Hamiltonian assumes the form of a two-component isotropic Ashkin-Teller model~\cite{baxter}  
\begin{equation}\label{AT}
H_{ex}=-J_1 \sum_{\langle i,j \rangle} (\tilde{s}_i \tilde{s}_j+\sigma_i\sigma_j)-J_2 \sum_{\langle i,j \rangle} \tilde{s}_i \tilde{s}_j \sigma_i \sigma_j.
\end{equation} 
The phase diagram of this model is shown in Fig.~\ref{ashkinteller} \cite{kadanoff-etal}, which we discuss below qualitatively in terms of the original variables $s$ and $\sigma$.

For weak Fermi surface anisotropy, which corresponds to small values of $J_{2}/J_{1} (\sim \delta^2$), there exists a continuous transition (across the dashed line in Fig.~\ref{ashkinteller}) from a high temperature disordered phase to a low temperature ordered phase. Along this line of direct transition between the disordered and the ordered phases, the exponents change \emph{continuously}, much like for the Kosterlitz-Thouless transition. In the ordered phase, the surface states break both translational (by the SDW order) and rotational (by the nematic order) symmetries, and the expectation values of the Ising-spin variables in Eq.~(\ref{exch}) are $\langle s \rangle \neq 0$ and $\langle \sigma \rangle \neq 0$. This phase is also known as the \emph{Baxter phase} \cite{baxter}. However, for large $\delta$ or $J_{2}/J_{1}$ (large Fermi surface anisotropy) transitions associated with these two symmetry breakings bifurcate and occur at distinct temperatures. The system first condenses into the nematic phase, where $\langle s \rangle \neq 0$ but $\langle s \sigma \rangle=\langle \sigma \rangle=0$, and only subsequently enters the ordered (Baxter) phase at lower temperature. Next we characterize each of these phases in terms of original order parameters, $\Delta_X$ and $\Delta_Y$.

The nematic phase is ordered along either ${\bf Q}_X=(\pi,0)$ or ${\bf Q}_Y=(0,\pi)$ in such a way that a large density of sign flips (domain wall) of the order parameter proliferate in the system. In this phase $ \langle|\Delta_X|\rangle$ or $ \langle |\Delta_Y| \rangle$ is non-zero, but $ \langle \Delta_X \rangle=\langle \Delta_Y \rangle=0$. Consequently, the nematic phase breaks the $C_4$ rotational symmetry, yet still retains the translational invariance of the the system. Only at lower temperature, through a subsequent transition system enters into the ordered/Baxter phase, where both nematic and density-wave orders develop non-zero expectation value. It is worth mentioning that a similar, but distinct, nematic phase has also been studied for iron-based superconductors \cite{fernandez-chubukov-schmalian, sachdev, Fernandes2}.

\section{Summary and Discussion}\label{sec:summary}

In summary, we discuss various many-body instabilities on the surface of strongly interacting cubic TKIs. We show that if the chemical potential is placed in between the Dirac points at the $\Gamma$ and $X/Y$ points of the surface BZ and the resulting electron (near $X$ and $Y$ points) and hole (near $\Gamma$ point) pockets are of comparable size, fermions can condense into a nematic and density-wave phase. In this phase only one of the electron pockets participates in excitonic pairing, and thus the 4-fold rotation symmetry on the surface gets lifted spontaneously. Therefore, our results provide a possible explanation for the recently observed $C_2$ symmetric magnetoresistence \cite{c4-magnetiresis} and the upturn in surface resistivity with tunable gate voltage or equivalently the chemical potential \cite{paglione} in SmB$_6$.

The excitonic phase, however, can display a spin- or charge-density wave ordering depending on the relative chirality of the Dirac cones with electron and hole like carriers. In Sec.~II, we argued that due to the presence of underlying strong spin-orbit coupling that causes spin(in-plane components)-momentum locking of the surface states \cite{review-1, review-2}, and only the $z$-component of electrons' spin participates in various instabilities, which in turn also allows the system to exhibit long-range order at finite-$T$. Our results are substantiated by complimentary Ginzburg-Landau analysis of order parameters (for small Fermi surface mismatch) in Sec. III and the free-energy minimization in mean-field approximation (for arbitrary values of the parameters $\mu, \delta, \lambda, T$) in Sec. IV. For large Fermi mismatch, on the other hand, our mean-field analysis predicts that both electron pockets gets involved in excitonic ordering, and the ordered phase restores the 4-fold rotational symmetry of the surface BZ.

Furthermore, we extend our analysis beyond the mean-field level, and account for thermal fluctuations and domain walls when the system condenses into a $C_2$ density-wave phase in Sec.~V. In this limit, the system can be described by a two-component isotropic Ashkin-Teller model and we presented a finite temperature phase diagram in Fig.~\ref{ashkinteller}. For small Fermi surface mismatch, both nematic and density-wave orders condense at the same temperature in agreement with our mean-field analysis. Only for substantial Fermi surface mismatch, these two transitions take place at different temperatures. System first pairs into a nematic phase and yet at a lower temperature to an excitonic (Baxter) phase. Although our study is primarily motivated by ongoing experimental works in SmB$_6$ \cite{c4-magnetiresis, paglione} that are suggestive of the presence of strong electronic correlations on its surface, it can describe various signature of electron-electron interactions on the surface of other strongly interacting cubic TIs, such as YbB$_6$ \cite{ybb6-1, ybb6-2, ybb6-3}, PuB$_6$ \cite{pub6}.

 Various recent experiments have extracted the effective parameters for the surface band structure. For example, ARPES experiments have found the ellipticity factor $\delta=0.1-0.4$~\cite{exp4, exp7, exp8, exp9}  not too large in SmB$_6$. Extracting the energy offset among the Dirac points in an experiment is a challenging task. Nevertheless, various first-principle~\cite{xidai} and effective band structure~\cite{roy-sau-dzero-galitski} calculations suggest that $|E_\Gamma-E_{X/Y}| \sim 2-10$ meV. The estimated values of these parameters indicates that while the ellipticity of the Fermi pockets is not too large to destroy the propensity of nematic and excitonic orderings on the surface, a large offset among the Dirac points allow one to tune the surface chemical potential over a reasonably wide range to realize electron and hole pockets of comparable sizes thourhg external gating~\cite{paglione}, conducive for orderings. Therefore, with currently estimated values of these band parameters, it is quite conceivable that surface states of SmB$_6$ or other cubic TKIs (such as YbB$_6$ and PuB$_6$) can accommodate various exotic broken symmetry phases.

 Detection of the nematic or the $C_2$ symmetric excitonic orderings demands direction dependent measurements of transport quantities, for example. Here, we focus only on the $C_2$ symmetry breaking ordering, as it occupies most of the phase diagram in Fig.~\ref{fig:phase0p2}. Notice in the nematic and the excitonic phases the four-fold rotational symmetry gets broken, while the former one is deviod of uniform condensation of any order parameter. Therefore, to pin the the onset of these orderings one needs to perform direction dependent measurements of various physical quantities, such as conductivity, resistivity, magnetoresistence, on the surface that can sense the lack of rotational symmetry in the close proximity to an ordering. Recent experiment~\cite{c4-magnetiresis} has reported the lack of four fold rotational symmetry in magnetoresistance in SmB$_6$ below 5K, which is suggestive of at least a nematic ordering on the surface.

It should be noted that the surface BZ of cubic TKI closely is similar to the one in pnictides~\cite{fernandez-chubukov-schmalian, sachdev, Fernandes2}. However, there exist several crucial differences between these two systems. For example, due to the strong spin-orbit coupling the SDW order of the surface states breaks only the discrete $Z_2$ symmetry [note that the valley $U(1)$ symmetry of SDW order is only an artifact of the low energy approximation in Eqs.~(\ref{freeHamil}), (\ref{eq:HSHamiltonian}) which gets reduced to $Z_2$ due to the presence of an underlying lattice captured by the term $\mathcal{F}_{SB}$ in Eq.~(\ref{SymBreaking})], responsible for true long-range order, whereas spin-rotation is a good symmetry and the SDW phase breaks continuous $SO(3)$ symmetry in pnictides \cite{sachdev}. In addition, the Fermi surfaces on the surface of TKIs constitute vortices or anti-vortices in momentum space that in turn encode the bulk topological invariance of the system, while the bands in pnictide materials are regular non-relativistic parabolic bands. Consequently, the regular parabolic bands and therefore the SDW order in pnictides can carry additional orbital degeneracy, which depends on various nonuniversal details of the system \cite{fernandez-chubukov-schmalian}, whereas the non-interacting model [see Eq.~(\ref{freeHamil})] and the SDW/CDW order [see Eq.(\ref{eq:HSHamiltonian})] we consider for the surface states TKIs is constrained by nontrivial bulk topological invariant. Thus, neither SDW nor CDW is accompanied by additional degeneracy. In additional, contrast to our results, a recent theoretical study finds that the transition from paramagnetic-$C_2$ SDW in pnictide is discontinuous or first order in nature \cite{chubukov-fernande-recent}. Despite these fundamental differences, we find that the qualitative structure of the phase diagram in Fig.~\ref{fig:phase0p2} for the surface states of TKIs bears some similarities to the one for iron pnictides both calculated theoretically \cite{fernandez-chubukov-schmalian} and also with the one obtained experimentally (see the phase digram of Ba$_{1-x}$Na$_x$Fe$_2$As$_2$ in Fig.~2 of Ref.\cite{Avci}). The similarity between such different systems is both surprising and encouraging. Therefore, we expect that our study will initiate future works related to TKIs that may unearth some exotic effects due to the presence of strong electronic correlations in these systems and may as well shed light into the phase diagram of iron pnictides.

As a final remark, we highlight some other possible phenomena, arising from strong residual electronic interactions on the surface of TKIs, among which the renormalization of plasmon spectrum due to strong fluctuations~\cite{efimkin-galitski}, non-Fermi liquid phase for $d$-electrons~\cite{nikolic}, quasi-particle inteference~\cite{vojta}, and spontaneous valley Hall ordering in the presence of strong magnetci field~\cite{li-roy}. In addition, a spatial variation of the hybridization has been proposed to lead to a topological chiral-liquid on the surface~\cite{onur}, without destroying the helical structure of the surface states (protected by bulk topological invariant). While these proposals are quite fascinating and of definite fundamental importance, our work focuses on the possibilities of realizing various broken symmetry phases (excitonic and nematic) on the surface of TKI, resulting from strong residual interactions, which can explain some peculiar experimental observations in recent past~\cite{c4-magnetiresis, paglione}.

\acknowledgements

We thank P.M.R. Brydon for useful discussions. B.R. and J.D.S. were supported by the startup grant of J.D.S by University of Maryland. J.H. was supported by LPS-CMTC. V.S. and V.G. were supported by DOE-BES DESC0001911, and Simons Foundation. B. R. is thankful to the Max-Planck Institute for Complex Systems, Dresden, Germany, for hospitality during the workshop ``Topology and Entanglement in Correlated Quantum Systems (2014)", where part of this work was completed. V. G. is grateful to Johnpierre Paglione and Michael Fuhrer for useful discussions.

\end{document}